\documentstyle[floats,prd,aps,epsfig,axodraw,eqsecnum,12pt]{revtex}

\makeatletter
\newbox\tempboxa
\newdimen\captionboxsubcount
\def\capsize#1{\captionboxsubcount=#1pt}
\newdimen\captionboxsub
\captionboxsub=\hsize \advance\captionboxsub by -\captionboxsubcount
\advance\captionboxsub by -\captionboxsubcount
\long
\def\@makecaption#1#2{
 \setbox\@tempboxa\hbox{#1 #2}
 \ifdim \wd\@tempboxa >\captionboxsub
\rightskip=\captionboxsubcount \leftskip=\captionboxsubcount #1 #2
\else \hbox to\hsize{\hfil\box\@tempboxa\hfil}
 \fi}
\makeatother
\capsize{30}

\begin{document}
\vfill
\begin{titlepage}
\begin{flushright}
\begin{minipage}{5cm}
\begin{flushleft}
\small
\baselineskip = 10pt
YCTP-P-02-00\\
\end{flushleft}
\end{minipage}
\end{flushright}

\begin{center}
\Large\bf
A Note on Anomaly Matching  for Finite Density QCD
\end{center}
\footnotesep = 12pt
\vskip 3cm
\begin{center}
\large
Francesco {\sc Sannino} \footnote{Electronic address : {\tt
francesco.sannino@yale.edu}}
\vskip 1cm
 {\it Department of Physics, Yale University, New
Haven,\\~CT~06520-8120,~USA.}
\end{center}
\vfill
\begin{center}
\bf Abstract
\end{center}
\begin{abstract}
\baselineskip = 17pt We note that the QCD phases at large finite
density respect 't Hooft anomaly matching conditions. Specifically
the spectrum of the light excitations possesses the correct quantum
numbers required to obey global anomaly constraints. We argue that
't Hooft constraints can be used at finite density along with non
perturbative methods to help selecting the correct phase.

\end{abstract}
\begin{flushleft}
\footnotesize
\end{flushleft}
\vfill
\end{titlepage}

\section{Introduction}

\label{uno}

Recently quark matter at very high density has attracted a great
flurry of interest \cite{ARW_98,RSSV_98,ARW_98b,SW_98b}. In this
regime quark matter is expected to behave as a color superconductor
\cite{ARW_98,RSSV_98}. Possible phenomenological applications are
associated with the description of neutron star interiors, neutron
star collisions and the physics near the core of collapsing stars.
A better understanding of highly squeezed nuclear matter might also
shed some light on nuclear matter at low density, i.e. densities
close to ordinary nuclear matter where some model already exist.
See for example Ref.~\cite{HSSW} where a rather complete soliton
model at low density is constructed containing along with the
Goldstone bosons also vector-bosons.

In a superconductive phase, the color symmetry is spontaneously
broken and a hierarchy of scales, for given chemical potential, is
generated. Indicating with $g$ the underlying coupling constant the
relevant scales are: the chemical potential $\mu$ itself, the
dynamically generated gluon mass $m_{gluon}\sim g \mu $ and the gap
parameter $\Delta\sim\mu e^{ -\frac{1}{g}}$. Since for high $\mu$
the coupling constant $g$ (evaluated at the fixed scale $\mu$) is
$\ll 1$, we have:
\begin{equation}
\Delta \ll m_{gluon} \ll \mu \ .
\end{equation}
Massless excitations dominate physical processes at very low energy
with respect to the gap ($\Delta$) energy. Their spectrum is
intimately related to the underlying global symmetries and the way
they are realized at low energies. Indeed when the dynamics is such
that a continuous global symmetry is spontaneously broken a
Goldstone boson appears in order to compensate for the breaking.
Massless excitations obey low energy theorems governing their
interactions which can be usefully encoded in effective
Lagrangians. A well known example, in the regime of cold and non
dense QCD, is the effective Lagrangian for pions and kaons. These
Lagrangians are seen to describe well the QCD low energy
phenomenology \cite{effective Lagrangians}.

Another set of relevant constraints is provided by quantum
anomalies. At zero density and temperature, 't Hooft \cite{thooft}
argued, using a beautiful mathematical construction, that the
underlying continuous global anomalies have to be matched in a
given low energy phase by a set of massless fermions associated
with the intact global symmetries and a set of massless Goldstone
bosons associated with the broken ones. The low energy fermions
(composite or elementary) contribute via triangle diagrams while
for Goldstones a Wess-Zumino term should be added to correctly
implement the associated global anomalies. In this note we
investigate the 't Hooft constraints for QCD at finite density.

In the next section we show, by reviewing the dynamically favored
phases for $N_f=2,3$ at high density that the low energy spectrum
displays the correct quantum numbers to saturate the 't Hooft
global anomalies.

We also observe that QCD at finite density can be envisioned, from
a global symmetry and anomaly point of view, as a chiral gauge
theory \cite{ball,ADS} for which at least part of the matter field
content is in complex representations of the gauge group. Indeed an
important distinction from, zero density, vector-like theories is
that these theories, when strongly coupled, can exist in the Higgs
phase by dynamically breaking their own gauge
symmetries\cite{ball,ADS}. This is also the striking feature of the
superconductive phase allowed for QCD at high density. In fact at
finite density vector like symmetries are no longer protected
against spontaneous breaking by the Vafa-Witten theorem \cite{VW}.

As for a chiral gauge and in general for any gauge theory we expect
the low energy massless spectrum of a finite density phase
(fermions associated to intact chiral global anomalies and
Goldstones for spontaneously broken symmetries) to possess the
quantum numbers required by the 't Hooft anomaly matching
conditions. This further global constraint, when appropriately
taken into account, can help selecting the low energy phase at
finite density.

\section{Anomaly matching at finite density}
We now show that the recently discussed superconductive QCD phases
at high density do respect 't Hooft anomaly matching conditions.
Specifically the spectrum of the light excitations possesses the
correct quantum numbers needed to satisfy global anomaly
constraints. The underlying gauge group is $SU(3)$ while the
quantum flavor group is
\begin{equation}
SU_L(N_f) \times SU_R(N_f) \times U_V(1) \ ,
\end{equation}
and the classical $U_A(1)$ symmetry is destroyed at the quantum
level by the Adler-Bell-Jackiw anomaly. We indicate with
$q_{\alpha;c,i}$ the two component left spinor where $\alpha=1,2$
is the spin index, $c=1,...,3$ is the color index while
$i=1,...,N_f$ represents the flavor. $\widetilde{q}^{\alpha ;c,j}$
is the two component conjugated right spinor. We summarize the
transformation properties in the following table.
\begin{equation}
\begin{tabular}{ccccc}
& $[SU(3)]$ & $SU_{L}(N_{f})$ & $SU_{R}(N_{f})$ & $U_{V}(1)$ \\ & &
&  &  \\ $q$ &
\begin{tabular}{|c|}
\hline
$\ \ $ \\ \hline
\end{tabular}
&
\begin{tabular}{|c|}
\hline
$\ \ $ \\ \hline
\end{tabular}
& $1$ & $1$ \\ &  &  &  &  \\ $\widetilde{q}$ & $\overline{
\begin{tabular}{|c|}
\hline
$\ \ $ \\ \hline
\end{tabular}
}$ & $1$ & $\overline{
\begin{tabular}{|c|}
\hline
$\ \ $ \\ \hline
\end{tabular}
}$ & $-1$ \ .
\end{tabular}
\end{equation}
The theory is subject to the following global anomalies:
\begin{equation}
SU_{L/R}(N_f)^3 \ , \quad SU_{L/R}(N_f)^2\, \times U_V(1) \ .
\end{equation}
For a vector like theory there are no further global anomalies. The
cubic anomaly factor, for fermions in fundamental representations,
is $1$ for $q$ and $-1$ for $\tilde{q}$ while the quadratic anomaly
factor is always $1$ leading to
\begin{equation}
SU_{L/R}(N_f)^3 \propto \pm 3 \ , \quad SU_{L/R}(N_f)^2 U_V(1)
\propto \pm 3 \ .
\end{equation}

We first consider the case $N_f=2$ which has only the
$SU_{L/R}(2)^2 \, \times U_V(1)$ anomaly. At zero density we have
two possible phases compatible with the anomaly conditions. The
first is the ordinary Goldstone phase associated with the
spontaneous breaking of the underlying global symmetry to
$SU_V(2)\times U_V(1)$. The other is the Wigner-Weyl phase where,
assuming confinement, the global symmetry at low energy is intact
and the needed massless spectrum consists of massless baryons with
the following quantum numbers:
\begin{equation}
\begin{tabular}{ccccc}
& $[SU(3)]$ & $SU_{L}(2)$ & $SU_{R}(2)$ & ${U}_{V}(1)$ \\ & & & &
\\ $B$ & $1$ &
\begin{tabular}{|c|}
\hline
$\ \ $ \\ \hline
\end{tabular}
& $1$ & $3$ \\ &  &  &  &  \\ $\widetilde{B}$ & $1$ & $1$ &
$\overline{%
\begin{tabular}{|c|}
\hline
$\ \ $ \\ \hline
\end{tabular}
}$ & $-3$%
\end{tabular}
\end{equation}
Here the two component baryon emerges as composite field of the
form
\begin{equation}
B_i=\epsilon^{abc} \epsilon^{jk} q_{a;i}q_{b,j} q_{c,k} \ ,
\end{equation}
where $a,b,c$ and $i,j,k$ are respectively color and flavor indices
and we have omitted spin indices. A similar expression holds for
$\tilde{B}$. The Goldstone phase is the one observed in nature.
This fact supports a new idea presented in Ref.~\cite{ADS}. Here it
is suggested that, for an asymptotically free theory\footnote{We
consider theories without flat directions. The possibility to
extend our criterion to these theories is presently under
investigation.}, among multiple infrared phases allowed by 't Hooft
anomaly conditions the one which minimizes the entropy at the
approach to freeze-out is preferred. The entropy $S(T)$ near freeze
out is given by $S(T)=(2\pi^{2}/45) T^{3} f_{IR}$
\cite{ADS,acs,acss} plus higher order terms in the low temperature
expansion. For the special case of an infrared-free theory,
$f_{IR}$ is simply the number of massless bosons plus $7/4$ times
the number of 2-component massless Weyl fermions. So the minimum
entropy guide at the freeze out can be viewed via $f_{IR}$ as a
minimum degree of freedom count. In the real world case of QCD with
two flavors the three Goldstone bosons lead to $ f_{IR}= 3$, and
the two massless composite fermions lead to $f_{IR}= 7$. Indeed, in
this case, nature chooses to minimize $S(T)$ at the approach to
freeze out
\footnote{Of course, at any finite T, the preferred phase is
chosen from among all the states entering the partition function by
minimizing the free energy density, which becomes the energy
density at $T=0$. Comparing these quantities for different states
when the theory is strongly interacting is, however, generally a
strong coupling problem.}.

This guide, although at a very speculative level, has been used in
\cite{ADS} to select the infrared phase of chiral gauge theories.

What happens to the 't Hooft anomaly conditions when  we squeeze
nuclear matter?  At very low baryon density compared to a fixed
intrinsic scale of the theory $\Lambda$, it is reasonable to expect
that the Goldstone phase persists. Hence 't Hooft anomaly
conditions are still satisfied. On the other hand at very large
densities it is seen, via dynamical calculations
\cite{ARW_98,RSSV_98}, that the ordinary Goldstone phase is no
longer favored compared with a superconductive one associated with
the following type of quark condensate:
\begin{equation}
\epsilon^{\alpha \beta} \epsilon^{abc} \epsilon^{ij}
< q_{\alpha;b,i}q_{\beta;c,j} > \ ,
\end{equation}
likewise for the tilded quarks. In the following we set $a=3$. This
condensate is not allowed at zero density by the Vafa-Witten
theorem. One can ask if the spectrum of low energy excitations
still possesses the correct quantum numbers to satisfy 't Hooft
anomaly conditions. The previous condensate breaks the gauge
symmetry while leaving intact the following group:
\begin{equation}
 \left[SU(2)\right]\times SU_L(2)\times SU_R(2)\times \widetilde{U}_V(1) \ ,
\end{equation}
where $\left[SU(2)\right]$ is the unbroken part of the gauge group.
The $\widetilde{U}_V(1)$ generator is the following linear
combination of the previous $U_V(1)$ generator $Q_V$ and the broken
diagonal generator of the $SU(3)$ gauge group $Q_{8}={\rm diag}
\{1,1,-2\}$
\begin{equation}
\widetilde{Q}=Q_V - Q_{8}\ ,
\end{equation}
Then the $\widetilde{Q}$ charge of the quarks with color $1$ and
$2$ is zero.

The low energy massless excitations are the quarks not
participating in the condensate. We summarize their symmetry
properties as follows:
\begin{equation}
\begin{tabular}{ccccc}
& $[SU(2)]$ & $SU_{L}(2)$ & $SU_{R}(2)$ & $\widetilde{U}_{V}(1)$ \\
&  &  &  &  \\
$q_{\alpha ;3,i}$ & $1$ &
\begin{tabular}{|c|}
\hline
$\ \ $ \\ \hline
\end{tabular}
& $1$ & $3$ \\
&  &  &  &  \\
$\widetilde{q}^{\alpha ;3,j}$ & $1$ & $1$ & $\overline{%
\begin{tabular}{|c|}
\hline
$\ \ $ \\ \hline
\end{tabular}
}$ & $-3$%
\end{tabular} \ ,
\end{equation}
These massless low energy fermions correctly match the 't Hooft
anomaly conditions. In Fig.~1 we draw the relevant diagrams for the
global anomalies. At low energies we use the elementary massless
fields $q_{3,i}~(\widetilde{q}^{3,i})$ with non zero
$\widetilde{Q}$ charge contributing to the first diagram. At high
energies we should consider the underlying symmetries and hence we
split the left hand side triangle into the two right hand side
diagrams and we exchange all of the underlying fermions. Since the
second diagram contains only one gauge generator it vanishes
identically and the 't Hooft anomaly conditions are matched.

Our proof is part of a more general theorem established for chiral
gauge theories at zero density. The theorem states that if in a
theory of massless fermions, one breaks the gauge symmetry in such
a way that a subset of the original fermions remain massless, those
massless fermions always obey 't Hooft anomaly conditions with
respect to the unbroken chiral symmetries \cite{peskin}.

The superconductive phase for $N_f=2$ possesses the same global
symmetry group of the confined Wigner-Weyl phase. This remarkable
feature when considering chiral gauge theories at zero density
(where the superconductive phase is now a Higgs phase) is referred
as complementarity. This idea was introduced in Ref.~\cite{rds}
where it was conjectured that any Higgs phase can be described in
terms of confined degrees of freedom and vice versa. We stress that
complementarity does not imply that an Higgs phase and a confined
phase are physically equivalent. Indeed the elementary or composite
nature of the low energy  particles can be uncovered via scattering
experiments. Hence, finite density QCD, at least from the global
symmetry and anomalies point of view, resembles a chiral gauge
theory.

However since the minimum entropy guide has been stated only at
zero density it should not be used to help selecting a phase at
high density. Clearly at low densities, where we expect the theory
to behave like ordinary QCD, the conjecture holds.

\begin{figure}
\vskip 2in
\centerline{
\begin{picture}(300,100)(-15,-15)
\put(-90,105){\makebox(0,0)[br]{$\tilde Q$}}
\Line(-60,110)(-10,150)
\Line(-60,110)(-10,60)
\Line(-10,60)(-10,150)
\Photon(-80,110)(-60,110)3 4
\Photon(-10,150)(10,150)3 4
\Photon(-10,60)(10,60)3 4
\put(60,150){\makebox(0,0)[br]{$SU_{L/R}(2)$}}
\put(60,60){\makebox(0,0)[br]{$SU_{L/R}(2)$}}
\put(45,105){\makebox(0,0)[br]{$=$}}
\put(75,105){\makebox(0,0)[br]{$Q_V$}}
\Line(100,110)(150,150)
\Line(100,110)(150,60)
\Line(150,60)(150,150)
\Photon(80,110)(100,110)3 4
\Photon(150,150)(170,150)3 4
\Photon(150,60)(170,60)3 4
\put(220,150){\makebox(0,0)[br]{$SU_{L/R}(2)$}}
\put(220,60){\makebox(0,0)[br]{$SU_{L/R}(2)$}}
\put(205,105){\makebox(0,0)[br]{$-$}}
\put(250,105){\makebox(0,0)[br]{$Q_{8}$}}
\Line(275,110)(325,150)
\Line(275,110)(325,60)
\Line(325,60)(325,150)
\Photon(250,110)(275,110)3 4
\Photon(325,150)(345,150)3 4
\Photon(325,60)(345,60)3 4
\put(395,150){\makebox(0,0)[br]{$SU_{L/R}(2)$}}
\put(395,60){\makebox(0,0)[br]{$SU_{L/R }(2)$}}
\end{picture}
}
\label{fig:1}
\caption{}
\end{figure}

Let us consider now the case of $N_f=3$ light flavors. At zero
density only the Goldstone phase is allowed and the resulting
symmetry group is $SU_V(3)\times U_V(1)$. Indeed there is no
solution of the 't Hooft anomaly condition with massless composite
fermions leaving intact the flavor group. In this case the
topological Wess-Zumino term for the Goldstone bosons can be
constructed to correctly implement the global anomalies of the
underlying theory at the effective Lagrangian level. The
Vafa-Witten theorem for vector-like theories prohibits the breaking
of vector like symmetries like $U_V(1)$.

Turning on low baryon density we expect to remain in the confined
phase with the same number of Goldstone bosons (i.e. 8). Evidently
the 't Hooft anomaly conditions are satisfied. At very high
density, dynamical computations suggest \cite{ARW_98b} that the
preferred phase is a superconductive one and the following ansatz
for a quark-quark type of condensate is energetically favored:
\begin{equation}
\epsilon^{\alpha \beta} < q_{\alpha;a,i}q_{\beta;b,j} > \sim k_1
\delta_{ai} \delta_{bj} + k_2 \delta_{aj} \delta_{bi} \ ,
\end{equation}
where we have a similar expression for the tilded fields. The
condensate breaks completely the gauge group while locking together
the left/right transformations to color. The final global symmetry
group is
\begin{equation}
SU_{c+L+R}(3) \ ,
\end{equation}
and the low energy spectrum consists of $9$ Goldstone bosons. The
effective Lagrangian at low energies \cite{CG} for the Goldstones
is similar to the ordinary effective Lagrangian for QCD at zero
density except for an extra Goldstone boson associated with the
spontaneously broken $U_V(1)$ symmetry. In this case we expect the
underlying global anomalies to be matched at low energies via the
Wess-Zumino term. It is instructive to explicitly construct this
term. This term is also been discussed together with its relation with
flavor anomalies at finite density in Ref.~\cite{HRZ}.
However in this paper, the two flavors QCD case is not investigated and the possibility to envision QCD at finite density as a chiral gauge theory is not discussed.

The Goldstone bosons are encoded in the unitary matrix $U$
transforming linearly under the left-right flavor rotations
\begin{equation}
U\rightarrow g_L U g_R^{\dagger} \ .
\end{equation}
with $g_{L/R} \in SU_{L/R}(N_f)$. In our notation $U$ is the
transpose of $\Sigma$ defined in Ref.~\cite{CG}. $U$ satisfies the
non linear realization constraint $UU^{\dagger}=1$. We also require
${\rm det} U=1$. In this way we avoid discussing the axial $U_A(1)$
anomaly at the effective lagrangian level. (see Ref.~\cite{SS} for
a general discussion of trace and $U_A(1)$ anomaly). We have
\begin{equation}
U=e^{i\frac{\Phi}{F^2}} \ ,
\end{equation}
with $\Phi=\sqrt{2}\Phi^a T^a$ representing the $8$ Goldstone
bosons. $T^{a}$ are the generators of $SU(3)$, with $a=1,...,8$ and
$\displaystyle{{\rm Tr}\left[ T^{a}T^{b}\right] =\frac{1}{2}\delta
^{ab}}$. $F$ is the Goldstone bosons decay constant at finite density.

The effective Lagrangian globally invariant under chiral rotations
is (up to two derivatives and counting $U$ as a dimensionless
field)
\begin{equation}
L=\frac{F^{2}}{2}\,{\rm Tr}\left[\partial_{\mu }U\partial^{\mu
}U^{\dagger }\right] \ .
\label{ef}
\end{equation}
The Wess-Zumino term \cite{WZ} can be compactly written using the
language of differential forms. It is useful to introduce the
algebra valued Maurer-Cartan one form:
\begin{equation}
\alpha=\left(\partial_{\mu}U\right)U^{-1}\, dx^{\mu}\equiv
\left(dU\right)U^{-1}\ .
 \label{MC}
\end{equation}
The Wess-Zumino effective action is
\begin{equation}
\Gamma_{WZ}\left[U\right]=C\, \int_{M^5} {\rm Tr} \left[\alpha^5\right] \ .
\label{WZ}
\end{equation}
The price to pay in order to make the action local is to augment by
one the space dimensions. Hence the integral must be performed over
a five-dimensional manifold whose boundary ($M^4$) is the ordinary
Minkowski space. The constant $C$, at zero density, is fixed to be
\begin{equation}
C=-i\frac{N}{240\pi^2} \ ,
\end{equation}
by comparing the current algebra prediction for the time honored
process $\pi^{0}\rightarrow 2\gamma$ with the amplitude predicted
using Eq.~(\ref{WZ}) once we gauge the electromagnetic sector
\cite{Witten,anomalies} of the Wess-Zumino term, and $N$ is the
number of colors (fixed to be 3 in this case). To the previous
lagrangian one can still add the extra Goldstone boson associated
to the $U_V(1)$ symmetry without altering the previous discussion
(see \cite{CG}). One can check that the global anomalies are
correctly implemented by carefully gauging the Wess-Zumino term
\cite{Witten,anomalies} with respect to the flavor symmetries.
Hence for the 3 flavor case too, the 't Hooft global anomalies are
matched at finite (low and high) density.

In writing the Goldstone Lagrangian we have not yet considered the
breaking of Lorentz invariance at finite density. Following
Ref.~\cite{CG} we note that the Goldstones obey, in medium, a
linear dispersion relation of the type $E={v}|{\vec{p}|}$, where
$E$ and $|{\vec{p}}|$ are respectively the energy and the momentum
of the Goldstone bosons. By rescaling the vector coordinates
$\vec{x}\rightarrow
\vec{x}/v$ the Lagrangian in Eq.~(\ref{ef}) becomes:
\begin{equation}
L=\frac{F^{2}}{2}\,{\rm Tr}\left[\dot{U}\dot{U}^{\dagger}
- v^2 {\vec{\nabla} U \cdot
\vec{\nabla}U^{\dagger}}\right]\ .
\end{equation}
$\alpha$, being a differential form, is unaffected by a coordinate
rescaling leaving unaltered the form of the Wess-Zumino term.

Due to the breaking of the baryon number the final global symmetry
group in the superconductive phase differs from the ordinary
Goldstone phase. Now we cannot regard this phase as complementary
to the confined one. As the baryon density decreases we also expect
the spectrum of light Goldstone bosons to change abruptly at the
phase transition.

Recently in Ref.~\cite{TS} QCD for $N_f$ larger than 3 has been
investigated at high density. All the phases discovered seem to
involve the breaking of global symmetries to a subgroup of the
vector like subgroup. These phases automatically respect 't Hooft
anomaly conditions.

\section{Conclusions}

 We noted, by reviewing the dynamically favored phases for
$N_f=2,3$ at high density that the low energy spectrum possesses
the correct quantum numbers to saturate the 't Hooft anomaly
matching conditions.

We have also argued that QCD at finite density can be thought, from
the point of view of global symmetry and anomalies, as a chiral
gauge theory \cite{ball}.  Indeed at finite density the vector like
symmetries are no longer protected against spontaneous breaking by
the Vafa-Witten theorem. As for a chiral gauge and in general for
any gauge theory we expect the low energy massless spectrum of a
finite density phase (fermions associated to intact chiral global
anomalies and Goldstones for spontaneously broken symmetries) to
possess the quantum numbers required by the 't Hooft anomaly
matching conditions. This further global constraint should be
appropriately taken into account when selecting the low energy
finite density phase.

We stress that, while dynamical calculations rely on the
high-density approximation, t'Hooft anomaly matching conditions are
global constraints and hence applicable to low as well as high
densities.

\vskip2cm \centerline{\bf Acknowledgments}
It is a pleasure for me to thank M.~Alford, R.L.~Jaffe and
K.~Rajagopal for enlightening discussions and encouragement. I
thank T. Appelquist for useful discussions and Z. Duan and P.S.
Rodrigues da Silva for helpful discussions as well as for careful
reading of the manuscript.
I thank I. Zahed for bringing to my attention some relevant literature.
A special thank goes to J. Schecther for
continuous support, encouragement and for a careful reading of the
manuscript. The work of F.S. has been partially supported by the US
DOE under contract DE-FG-02-92ER-40704.

\end{document}